\documentclass[a4,10pt,onecolumn]{article}
\usepackage{graphicx} % Required for inserting images
\usepackage[margin=0.8in]{geometry} 
\usepackage{amsmath,amsthm,amssymb}
\usepackage{float}
\usepackage{xcolor}
\usepackage[utf8]{inputenc}
\usepackage{graphicx}
\usepackage[backend=bibtex,style=nature]{biblatex}
\usepackage{authblk}

\def\_#1{{\bf #1\mit}}

\graphicspath{{Figures_temp/}{Figures_final/}}
\usepackage[font=footnotesize]{caption}

\title{Novel functionalities enabled by asymmetric reconfiguration in VO2-dielectric metasurfaces}

\date{}

\author[1]{P. Hostalet-Vicent$^*$}
\author[1]{L.M. Máñez-Espina$^*$}
\author[2]{K. Schouteden}
\author[2]{ J. P. Locquet}
\author[2]{M. Seo}
\author[1]{T. Mengual-Chulia}
\author[1]{ P. Sanchis}
\author[1]{ A. Díaz-Rubio}
\affil[1]{Nanophotonics Technology Center, Universitat Politècnica de València, València 46022, Spain.}
\affil[2]{Department of Materials Engineering, KU Leuven, Kasteelpark Arenberg 44, Leuven, Belgium}

\begin{document}

\maketitle

$^*$These authors have contributed equally

\begin{abstract} Bianisotropic metasurfaces provide electromagnetic functionalities that cannot be achieved through electric and magnetic responses alone, enabling asymmetric scattering, directional absorption, and advanced wavefront control. At optical frequencies, however, most demonstrations rely on static nanostructures whose magnetoelectric response is fixed after fabrication. Here, we investigate the opportunities enabled by dynamic reconfiguration of electromagnetic bianisotropy using a thermally tunable silicon--VO$_2$ metasurface operating in the near-infrared. By combining a collective-polarizability framework with numerical simulations and experimental characterization, we show that the insulator-to-metal transition of VO$_2$ modifies not only the electric and magnetic responses of the metasurface, but also its magnetoelectric coupling, providing access to distinct bianisotropic states within a single platform. Temperature-dependent experimental characterization confirms the thermally driven evolution of the absorption asymmetry. Based on this framework, we identify different operational regimes and demonstrate how bianisotropy reconfiguration enables transmissive and reflective directional functionalities, including transmission power-limiting, power-dependent asymmetric reflection, and power-dependent absorptivity/emissivity. These results establish dynamic control of magnetoelectric coupling as a powerful degree of freedom for multifunctional and adaptive photonic metasurfaces for optical and thermal applications. \end{abstract}
%\linenumbers.

\section{Introduction}
Metasurfaces have revolutionized wavefront engineering by enabling precise control of the amplitude, phase, and polarization of light within ultrathin platforms~\cite{Yu2011,Yu2014, khorasaninejad2016metalenses,park2017dynamic,quevedo2019roadmap}. While the majority of designs are based on electric and magnetic resonances, including Huygens' metasurfaces that provide efficient manipulation of electromagnetic waves, several fundamental wave transformations remain inaccessible when these responses act independently~\cite{Pfeiffer2013, pfeiffer2014bianisotropic,asadchy2018bianisotropic}. Bianisotropic metasurfaces overcome this limitation through magnetoelectric coupling, whereby electric and magnetic dipole moments become intrinsically linked, introducing additional degrees of freedom for electromagnetic field control~\cite{tretyakov1998magnetoelectric}. This enhanced design space enables functionalities such as asymmetric scattering, independent control of reflection and transmission under opposite illumination directions, advanced polarization manipulation, and high-efficiency wavefront shaping~\cite{asadchy2018bianisotropic}. Importantly, bianisotropy has been shown to be a necessary ingredient for realizing certain canonical transformations, including perfect anomalous refraction and reflection without parasitic diffraction while maintaining impedance matching~\cite{tretyakov2015metasurfaces,asadchy2016perfect}. Despite these remarkable capabilities, most bianisotropic metasurfaces demonstrated at optical frequencies rely on static nanostructures whose magnetoelectric response is entirely determined during the design and fabrication stages~\cite{shaltout2019spatiotemporal,gu2023reconfigurable}. As a result, their functionality remains fixed after fabrication, limiting the realization of dynamically reconfigurable photonic platforms capable of adapting their bianisotropic response in real time.
%These unique capabilities position bianisotropic metasurfaces as a powerful platform for next-generation flat optics and multifunctional photonic devices.

To overcome the static nature of conventional metasurfaces, a wide range of reconfiguration mechanisms have been explored to dynamically modify their optical response~\cite{gu2023reconfigurable}. Among them, phase-change materials (PCMs) have emerged as a particularly attractive platform owing to their large and reversible refractive-index modulation, enabling substantial tuning of metasurface functionalities within compact footprints~\cite{wuttig2017phase,kim2019phase}. This mechanism has enabled dynamic control of beam steering~\cite{kaplan2015dynamically,shabanpour2020ultrafast}, optical limiting~\cite{howes2020limiting,wan2021limiter}, tunable filtering~\cite{tripathi2021tunable}, adaptive optics~\cite{kim2019phase}, holography~\cite{wang2016optically}, and computational imaging~\cite{cotrufo2024reconfigurable}, highlighting the potential of reconfigurable metasurfaces for next-generation photonic systems. %Beyond reconfiguring conventional resonant responses, the large optical contrast provided by PCMs offers an opportunity to dynamically tailor the magnetoelectric coupling itself, opening the possibility of accessing fundamentally different bianisotropic regimes within a single metasurface platform.
%that cannot be described solely in terms of the initial or final optical response of the device.
When combined with PCMs, the implications of bianisotropy extend beyond conventional active tuning of the optical response. The phase transition can modify not only the electric and magnetic responses of a metasurface, but also its magnetoelectric coupling~\cite{manez2024control}, providing access to different bianisotropic states within a single platform. This additional degree of freedom substantially enlarges the available design space and enables directional functionalities. In particular, the interplay between asymmetric absorption and thermally driven phase transitions has recently enabled self-regulated directional functionalities, including nonreciprocal optical transmission without external biasing~\cite{king2024electrically,tripathi_nanoscale_2024}. 

Despite these promising demonstrations, the broader landscape of functionalities enabled by reconfigurable bianisotropy remains unexplored. A more general understanding requires a framework capable of relating the evolution of the bianisotropic response to the resulting scattering properties. In this context, theoretical descriptions based on resonant coupling, characterized by collective polarizabilities, provide a natural route to connect the underlying electromagnetic interactions to the macroscopic optical response of the metasurface~\cite{Serdyukov2001, Luis2026strong}. Such an approach offers a systematic way to identify the possibilities and the fundamental limitations associated with bianisotropy reconfiguration.

To explore the implications of bianisotropy reconfiguration in a realistic physical system, we consider a specific metasurface design that serves as a representative platform for the analysis. Among the available PCMs, we select VO$_2$ as it is particularly attractive for thermally reconfigurable photonic devices owing to its reversible insulator-to-metal transition. This phase transition produces a pronounced modification of the optical properties while operating close to room temperature~\cite{cavalleri2001femtosecond,shao2018recent,kim2019phase,cueff2020vo2}. In parallel, silicon nanodisk metasurfaces integrated with VO$_2$ have been investigated as tunable resonant platforms, establishing a foundation based on the understanding of the electric and magnetic dipolar resonances that govern their optical response~\cite{howes2020limiting,kepic2021tunablemie,tripathi2021tunable,Kang:22,oguntoye2023continuously}. Despite some applications studied in terms of bianisotropy and reconfiguration being reported~\cite{tripathi_nanoscale_2024}, a comprehensive study of the different electromagnetic responses that Omega bianisotropic metasurfaces using PCMs can allow is still missing. 

In this work, we make use of the collective-polarizability description of a thermally reconfigurable silicon-VO$_2$ metasurface to investigate the electromagnetic functionalities enabled by Omega bianisotropy reconfiguration. Through theoretical analysis supported by experimental characterization, we identify different operational regimes and illustrate how reconfigurable bianisotropy can be exploited to realize distinct directional power-dependent optical functionalities within a single metasurface platform.

%\textcolor{magenta}{Fig1(a), para explicar la funcionalidad a alto nivel}
%Here, we demonstrate a dynamically reconfigurable infrared bianisotropic metasurface enabled by the phase transition of VO\textsubscript{2}. By actively modulating the magnetoelectric coupling between the electric and magnetic resonances, the metasurface can be switched between fundamentally different absorption regimes, exhibiting either symmetric or strongly asymmetric absorption under opposite illumination directions. 

\section{Results and discussions}

\subsection{Optical description}
Bianisotropic metasurfaces can be conveniently described through a homogenized model based on collective polarizabilities, which relate the induced electric and magnetic dipole moments to the incident electromagnetic fields. For subwavelength periodic metasurfaces under normal incidence illumination, the induced electric and magnetic polarization densities can be expressed as~\cite{Albooyeh2023}

\begin{equation} \mathbf{P} = \frac{\overline{\overline{\alpha}}^{\,\mathrm{ee}}}{S}\cdot\mathbf{E}_{\mathrm{i}} + \frac{\overline{\overline{\alpha}}^{\,\mathrm{em}}}{S}\cdot\mathbf{H}_{\mathrm{i}}, \label{P}\end{equation} \begin{equation} \mathbf{M} = \frac{\overline{\overline{\alpha}}^{\,\mathrm{me}}}{S}\cdot\mathbf{E}_{\mathrm{i}} + \frac{\overline{\overline{\alpha}}^{\,\mathrm{mm}}}{S}\cdot\mathbf{H}_{\mathrm{i}}, \label{M} \end{equation} where $\mathbf{P}$ and $\mathbf{M}$ denote the induced electric and magnetic surface polarization densities, respectively, and $S$ is the unit-cell area. Within this formulation, $\overline{\overline{\alpha}}^{\mathrm{ee}}$ and $\overline{\overline{\alpha}}^{\mathrm{mm}}$ are tensors that describe the collective electric and magnetic polarizabilities, while $\overline{\overline{\alpha}}^{\mathrm{em}}$ and $\overline{\overline{\alpha}}^{\mathrm{me}}$ account for the magnetoelectric coupling responsible for bianisotropy. This description provides a direct connection between the underlying electromagnetic interactions and the resulting optical response of the metasurface.

In this work, we will restrict the study to Omega bianisotropic metasurfaces, which present antisymmetric off-diagonal elements in their polarizability tensors. Moreover, uniaxial symmetry will be imposed. In particular, the structure will present $C_{4v}$ point group symmetry. Considering only the zeroth diffraction order channel and the symmetries of the system, the transmission and reflection coefficients can be directly expressed in terms of the collective polarizabilities as

\begin{equation}\label{eq: t} t = 1-\frac{jk}{2S} \left( \eta {\alpha}^{\mathrm{ee}} + \frac{{\alpha}^{\mathrm{mm}}}{\eta} \right), \end{equation} \begin{equation}\label{eq: r} r^{\mathrm{f,b}}= -\frac{jk}{2S} \left( \eta {\alpha}^{\mathrm{ee}} - \frac{{\alpha}^{\mathrm{mm}}}{\eta} \pm 2{\alpha}^{\mathrm{em}} \right), \end{equation}
%\textcolor{orange}{revisar signo, en \cite{Luis2026strong} está así pero en \cite{manez2024control} está al revés. comentar el paso de tensores a elidir los 'hat'  y los índices xx,xy,yy }
where f and b denote forward and backward illumination, respectively. Here, $\alpha^{\rm em}=\alpha^{\rm em}_{xy}=-\alpha^{\rm em}_{yx}$, $\alpha^{\rm ee}=\alpha^{\rm ee}_{xx}=\alpha^{\rm ee}_{yy}$, and $\alpha^{\rm mm}=\alpha^{\rm mm}_{xx}=\alpha^{\rm mm}_{yy}$ are the only nonzero components of the polarizability tensors. The subscripts in these quantities denote the spatial directions of the tensorial components. Eqs.~\eqref{eq: r}-~\eqref{eq: t} reveal the different roles of the three polarizability components. While the transmission coefficient is governed solely by the balance between the electric and magnetic responses, the magnetoelectric polarizability contributes exclusively to reflection, introducing a directional dependence in the scattering response. Consequently, reciprocal omega-type bianisotropic metasurfaces preserve identical transmission for both illumination directions while exhibiting different reflection and absorption.

\begin{figure}[t]
    \centering
    \includegraphics[width=1\linewidth]{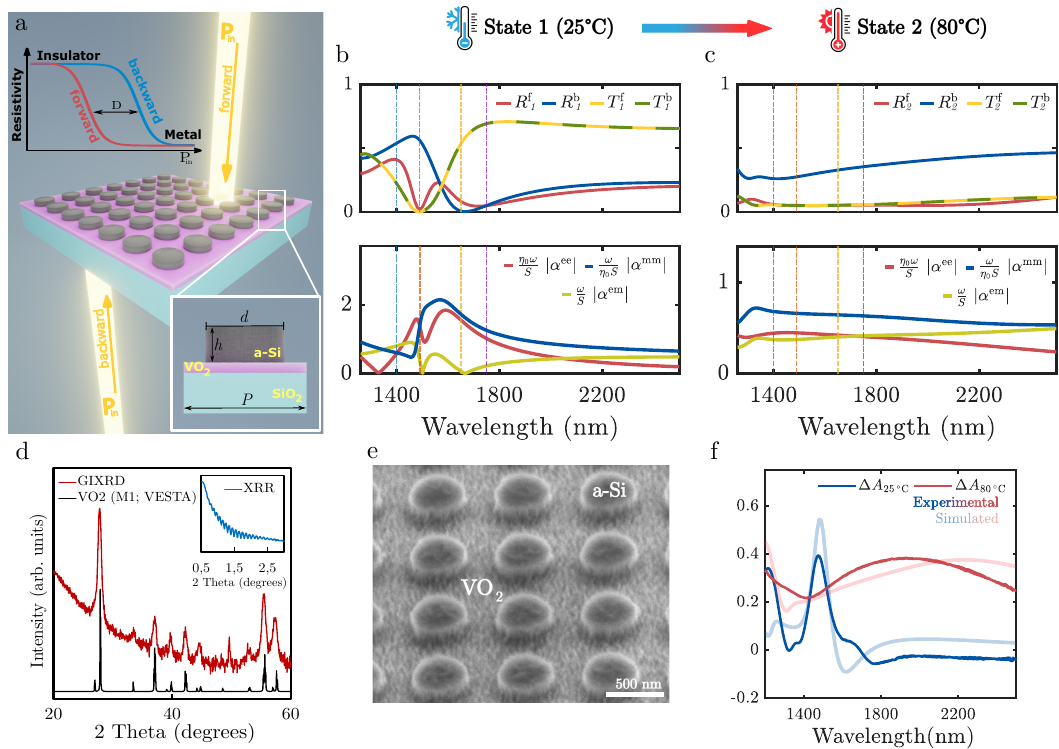}
       \caption{\textbf{Optical description of the thermally reconfigurable bianisotropic metasurface.} (a) Schematic of the proposed metasurface. Forward and backward illumination directions are defined in the image. (b-c) Simulated optical response and retrieved collective polarizabilities of the metasurface at 25$^\circ$C (b) and 80$^\circ$C (c), corresponding to the insulating and metallic states of VO$_2$, respectively. The thermally induced modification of the scattering properties results in a wavelength-dependent reconfiguration of the bianisotropic response, providing access to the different functional regimes explored throughout this work. The dashed lines indicate the wavelengths: $1400$, $1490$, $1650$, and $1750$nm for consideration in Section 3. The geometrical parameters of the simulated structure are $d=550$ nm, $h=228$ nm, $P=825$ nm, and VO$_2$ thickness of 77 nm. (d) Grazing Incidence X-Ray Diffraction (GIXRD) measurements confirming the monoclinic crystalline structure of VO$_2$ at room temperature. The thickness of the VO$_2$ layer (77 nm) is obtained by fitting the X-Ray Reflectometry (XRR) measurements. (e) Scanning electron microscopy (SEM) image of the lithographed silicon nanodisk pattern. (f) Absorption asymmetry, $\Delta A$, extracted from experimentally measured scattering parameters (dark lines) and compared with numerical simulations (light lines).} \label{Fig1}
\end{figure}

Several relevant electromagnetic regimes can be readily identified from these expressions. When the electric and magnetic responses satisfy the Huygens condition, $\eta{\alpha}^{\mathrm{ee}} = -{\alpha}^{\mathrm{mm}}/\eta$, and in the absence of magnetoelectric coupling, ${\alpha}^{\mathrm{em}} =0$,  reflection is strongly suppressed, and the metasurface approaches unity transmission~\cite{Pfeiffer2013,Cuesta2018}. Another regime can be achieved when a suitable balance between the electric, magnetic, and magnetoelectric responses leads to the balanced-particle condition, $\lvert{\eta{\alpha}^{\mathrm{ee}}}\rvert  = \lvert{\alpha}^{\mathrm{mm}}/\eta\rvert = \lvert {\alpha}^{\mathrm{em}}\rvert$, which maximizes the difference between forward and backward scattering. Indeed, with a structure presenting only Omega-type bianisotropy, in the presence of lossy materials, reciprocity ensures that the only independent scattering coefficients are $r^{\rm f}$, $r^{\rm b}$, and $t$. Here, the uniaxial symmetry of the structure grants it polarization-insensitive scattering. Furthermore, absorption can be computed by simply $A^{\rm f}=1-R^{\rm f}-T$ and $A^{\rm b}=1-R^{\rm b}-T$, where we have used capital letters for reflectance and transmittance, $R^{\rm f, b}=|r^{\rm f, b}|^2$ and $T=|t|^2$. Additionally, the difference in absorption between opposite illuminations is defined as $\Delta A=A^{\rm f}-A^{\rm b}=R^{\rm b}-R^{\rm f}$, i.e., the difference in absorption is only due to a difference in reflectance.

As a platform to investigate reconfigurable Omega metasurfaces, we consider the thermally reconfigurable metasurface shown in Figure~\ref{Fig1}(a), consisting of a square array of period $P$ conformed by amorphous silicon nanodisks with height $h$ and diameter $d$ on a VO$_2$ layer supported by a SiO$_2$ substrate. The out-of-plane asymmetry of the structure is introduced by the VO$_2$ layer and the SiO$_2$ substrate. This asymmetry can be controlled by the thermal reconfiguration of the VO$_2$ layer. The refractive index change in the layer will affect the bianisotropic term and therefore change the reflectance and absorption from the different sides of the structure. Moreover, since optical absorption is the driving phenomenon for the heating of the VO$_2$ layer, and it can differ between sides, the thermally induced phase transition can occur at different incident intensities depending on the illumination direction. More precisely, the thermal reconfiguration of the metasurface is driven by the insulator-to-metal transition of VO$_2$.

To achieve high contrast in absorption between sides, at normal incidence, we design the metasurface to support two dipolar Mie-type resonances. With a dielectric dipole resonance and an electric dipole resonance that, due to the VO$_2$ layer and the Silica substrate, are hybridized, i.e., present Omega bianisotropic behavior. For the proposed implementation, the geometrical parameters were selected as $d=550$ nm, $h=228$ nm, and $P=825$ nm, to approach this regime while simultaneously operating near the balanced-particle condition at 1490 nm. Indeed, at room temperature, the electric and magnetic resonances are engineered to be sufficiently close to promote strong hybridization and absorption. In particular, the resonance wavelengths for the dipolar resonances are 1482 nm for the hybrid electric-dominant and 1568 nm for the hybrid magnetic-dominant. The hybridization of the modes is also shown in their field profiles, see Supplementary Section 1. 
%For clarity, the optical response will be described in terms of two representative states corresponding to the insulating and metallic phases of the material, hereafter referred to as State 1 (25$^\circ$C) and State 2 (80$^\circ$C), respectively.

The simulated optical response and the retrieved collective polarizabilities are presented in Figures~\ref{Fig1}(b) and~\ref{Fig1}(c) for the insulator, State 1, and the metallic phase, State 2, respectively. As expected from Eq.~(4), the spectral evolution of the absorption asymmetry closely follows that of the magnetoelectric polarizability. Moreover, at $25^{\circ}\mathrm{C}$, see Figure~\ref{Fig1}(b), due to the lossy character of VO$_2$ and the low radiative quality factors, the resonances have comparable decay rates due to intrinsic losses and scattering. Such a condition makes possible absorption asymmetry with a pronounced resonant behavior in the region in which the two dipole-like resonances interact, around 1490 nm~\cite{Luis2026strong}. From the polarizabilities in Figure~\ref{Fig1}(b), we can see that maximum asymmetry in absorption, maximum difference between reflectances, is reached around the spectral region where the metasurface approaches the balanced-particle condition. Additionally, in this insulator state, for longer wavelengths, 1700 nm and beyond, the resonant behavior is lost, and the structure response is primarily due to the bulk layers of $\mathrm{VO}_2$ and the $\mathrm{SiO}_2$.

In the second state, after the VO$_2$ transition, which takes place at an approximate temperature of $68^{\circ}\mathrm{C}$~\cite{zouini2023literature}, the dielectric dipole resonances shift towards shorter wavelengths, around 1300 nm, see Figure \ref{Fig1}(c). This is due to the real part of the refractive index of the VO$_2$ decreasing in that wavelength region, see Supplementary Section 3. Moreover, the resonances become weaker, almost vanishing, due to the considerable increase in the imaginary part of the refractive index of VO$_2$ in the metallic phase. This increase makes decay rates due to intrinsic losses much higher than due to scattering, which in turn decreases the absorption in the system and resonance depth~\cite{Jose2023}. Moreover, at this state, the metallic phase of VO$_2$ shows an epsilon-near-zero behavior at wavelengths around 1170 nm, see Supplementary Section 2, that enhances absorption in the VO2 layer at shorter wavelengths. At longer wavelengths, the asymmetry in absorption is promoted by the metasurface on top of the VO2 layer. The metasurface smoothes the refractive index transition between air and the absorbing layer for the forward illumination, acting as an impedance matching layer.

%The selected wavelengths highlighted in Figures~\ref{Fig1}(b)-(c) correspond to different scenarios in which the ratio between electric, magnetic, and magnetoelectric polarizabilities enables different power-dependent scenarios. This will be later exploited in Section~\ref{Section: Functionalities} of this work.
%The retrieved polarizabilities shown in Figure \ref{Fig1}(b) reveal that the balance between the electric and magnetic responses is approximately maintained at the remaining wavelengths selected as representative examples throughout this work, namely 1400, 1650 and 1750 nm

The simulated metasurface was fabricated on VO$_2$ films grown on fused-silica substrates (see Methods). The crystalline phase and thickness of the active layer were verified by GIXRD and XRR measurements, respectively, confirming the formation of monoclinic VO$_2$ and a thickness of 77 nm (Figure \ref{Fig1}(d)). In addition, temperature-dependent resistivity measurements revealed the characteristic hysteretic insulator-to-metal transition of VO$_2$, confirming the reversible reconfiguration of the active layer (Supplementary Section 2). Subsequently, the silicon nanodisks were patterned to form the metasurface, as confirmed by the scanning electron microscopy image shown in Figure~\ref{Fig1}(e).

The optical response of the fabricated metasurface was characterized by FTIR measurements under forward and backward illumination at room temperature and above the VO$_2$ transition temperature. From the measured transmission and reflection spectra, the absorption asymmetry, $\Delta A$, was extracted and compared with numerical simulations, as shown in Figure~\ref{Fig1} (f). A good agreement is observed for both phase states of VO$_2$, reproducing the main spectral features predicted by simulations. Transmittance and reflectance measurements are provided in Supplementary Section 3. From Figure~\ref{Fig1} (f), it can be seen how the absorption asymmetry peaks at the resonant wavelength of around 1490 nm in the insulator state of the VO2. Moreover, a second resonant feature in the vicinity of 1600nm arises with opposite sign. This feature stems from mode hybridization, more precisely, electromagnetic mode coupling, due to the VO2 substrate breaking out-of-plane mirror symmetry~\cite{Luis2026strong}. In the metallic state of VO2, the metasurface exhibits positive absorption asymmetry across the entire spectrum. The presented measurements experimentally confirm that the bianisotropic response of the metasurface can be thermally reconfigured, providing the basis for the different electromagnetic functionalities analyzed in this work.

\subsection{\label{Section: Functionalities}Electromagnetic functionalities enabled by bianisotropy reconfiguration}

\begin{figure}[t]
    \centering
    \includegraphics[width=1.0\linewidth]{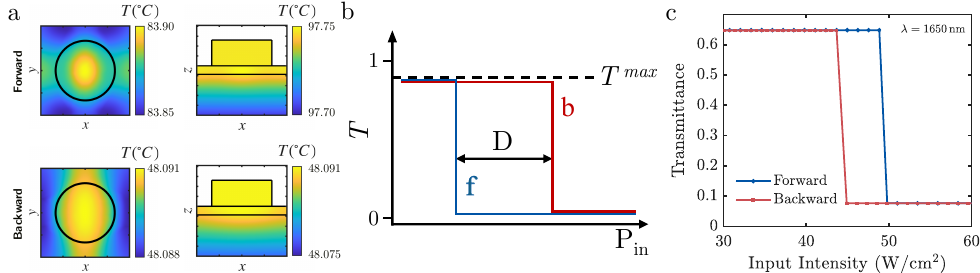}
      \caption{\textbf{Transmissive functionalities enabled by bianisotropy reconfiguration.} (a) Steady-state temperature distributions under forward and backward illumination at an incident intensity of 25~W/cm$^2$, obtained from the thermo-optical simulations. Cross-sectional and in-plane views show that optical absorption generates localized heating patterns following the electromagnetic field distribution. (b) Schematic representation of the optical isolation functionality. The absorption asymmetry of the initial state produces different phase-transition thresholds under forward and backward illumination, creating a range of incident powers where the transmission becomes direction dependent. (c) Simulated transmission of the proposed metasurface at 1650 nm as a function of incident optical intensity, illustrating the optical isolation regime arising from the asymmetric thermally induced transition.}
    
    \label{Fig3}
\end{figure}

In this section, we present the regimes of interest that arise when Omega bianisotropy is harnessed in conjunction with phase-change materials. The highlighted wavelengths shown in Figure~\ref{Fig1}(b) and (c) are the spectral points in which our analysis will focus, as they show distinct operational regimes for the response of the metasurface as a function of input power. 

The power-dependent response was obtained using COMSOL Multiphyisics from steady-state thermo-optical simulations following the procedure described in Methods. As illustrated in Figure~\ref{Fig3}(a), the resulting temperature distribution is largely determined by the heat dissipation conditions, leading to relatively small spatial variations across the active layer despite the strongly localized optical resonances. Nevertheless, the in-plane temperature profile reveals that the generated heat follows the local electromagnetic field distribution, as expected from the considered homogeneous complex refractive index of VO$_2$ within each phase state. To characterize the phase transition, the effective temperature of the metasurface is defined as the average temperature along the bottom interface of the VO$_2$ layer, which provides a robust indicator of the thermal state governing the optical response. 

The first application of interest, power limiting in transmission, has been covered extensively and experimentally proven to work even at low input intensities~\cite{howes2020limiting,tripathi_nanoscale_2024}. An initial $\Delta A_1 \neq 0$, results in different phase-transition thresholds under forward and backward illumination, as schematically illustrated in Figure~\ref{Fig3}(b). Consequently, the metasurface reaches the thermally reconfigured state at different incident intensities depending on the propagation direction, creating a finite power range where the device would behave differently if illuminated from opposite sides, one at a time. This intermediate regime gives rise to direction-dependent transmission limiting, with a window of operation quantified by $D$.

The operation of this regime is intrinsically constrained by the same bianisotropic response that drives the asymmetric transition. We can see from the scattering expressions introduced in \eqref{eq: t}-\eqref{eq: r}, that a non-zero absorption asymmetry requires some magnetoelectric coupling, which inevitably contributes to reflection and absorption under at least one illumination direction. As a result, increasing $\Delta A_1$ enhances the separation between transition thresholds and therefore enlarges $D$, but simultaneously reduces the maximum transmission that can be attained in the initial state. Consequently, nonreciprocal(asymmetric) transmission-limiting functionalities enabled by bianisotropy reconfiguration are governed by a fundamental trade-off between the achievable operation window and the transmission contrast between opposite illuminations.

In our implemented metasurface, we identify wavelengths for which the initial state remains close to the Huygens condition, $\eta{\alpha}^{\mathrm{ee}}\approx -{\alpha}^{\mathrm{mm}}/\eta$, while preserving a finite magnetoelectric coupling. Under these ideal conditions, the non-transmitted portion of the incident light is totally absorbed for one propagation direction while being completely reflected for the opposite one, maximizing the achievable transmission for a given threshold separation $D$. For the proposed metasurface, this regime is obtained around $\lambda=1650$~nm, as shown in Figure~\ref{Fig3}(c). Under these conditions, the incident intensity required for reconfiguration is on the order of $50~\mathrm{W/cm^2}$. This value is comparable to those previously reported in both numerical and experimental demonstrations employing continuous-wave excitation and thinner VO$_2$ layers~\cite{tripathi_nanoscale_2024}. We note that our thermo-optical model evaluates heat generation directly from the local optical absorption determined by the near-field distribution, providing a spatially resolved description of the reconfiguration process.

More generally, transmission asymmetry could be further enhanced by reducing ${\alpha}^{\mathrm{em}}$ while maintaining the overlap between the electric and magnetic resonances. This could be achieved through a more symmetric design, for example, by modifying the VO$_2$ layer position or by embedding the metasurface in a dielectric superstrate such as PMMA, as shown in \cite{tripathi_nanoscale_2024}.

%In this functionality, the bianisotropy of State~2 plays only a secondary role, since after reconfiguration, only $T_2^f \approx0$  is sufficient to enable the isolation regime. Transmissive operation does not benefit from arbitrarily increasing the magnetoelectric coupling. Reflective functionalities, in contrast, are not constrained by the same transmission-efficiency trade-off, and therefore able to exploit bianisotropy reconfiguration directly.

In the case of reflectance/absorptance reconfigurability, depending on the evolution of the absorption asymmetry as a function of the input power, three representative regimes can be identified:$(\Delta A_1>0,\Delta A_2>0)$, $(\Delta A_1>0,\Delta A_2\approx0)$, and $(\Delta A_1\approx0,\Delta A_2>0)$. Although, in principle, both $\Delta A_1$ and $\Delta A_2$ can take any value between $-1$ and $1$, the evolution of the collective polarizabilities in the proposed metasurface restricts the accessible reconfiguration paths to the positive one. However, notice how changing the signs of $\Delta A$ can be achieved by flipping the device itself. 

%We analyze representative points across the spectrum corresponding to the three functionalities shown in Figure~\ref{Fig4}.

\begin{figure}[h]
    \centering
    \includegraphics[width=1.0\linewidth]{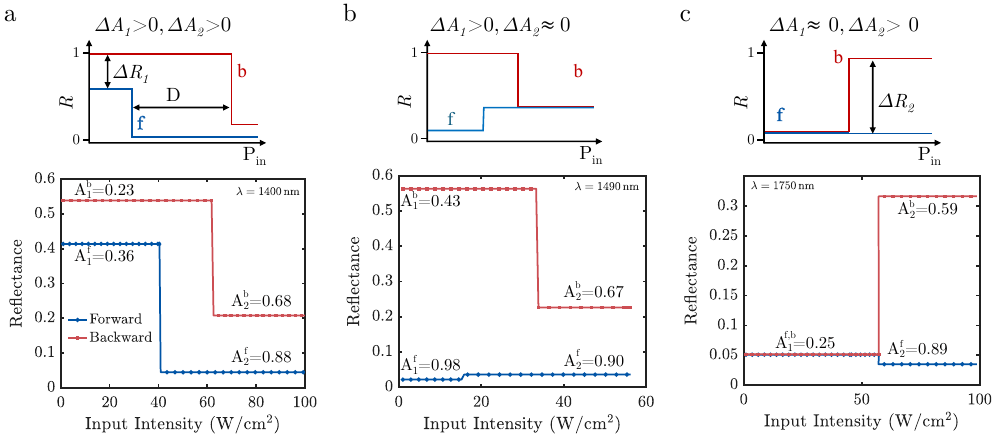}
       \caption{\textbf{Reflective functionalities enabled by bianisotropy reconfiguration.} For each functionality, the upper panels show a schematic representation of the operating regime, while the lower panels present the simulated response of the proposed metasurface. (a) Bi-directional asymmetric reflection. A non-zero absorption asymmetry in both states of the metasurface ($\Delta A_1>0$, $\Delta A_2>0$) results in asymmetric reflection before and after the thermally induced transition. Simulated at $\lambda=1400$ nm. (b) Directional reflective limiter. The metasurface evolves from a highly asymmetric state towards a nearly symmetric response ($\Delta A_1>0$, $\Delta A_2\approx0$). Simulated at $\lambda=1490$ nm. (c) Directional reflective switch. A nearly symmetric initial state evolves into a strongly asymmetric reflective response after thermal reconfiguration ($\Delta A_1\approx0$, $\Delta A_2>0$), resulting in a pronounced directional reflectivity contrast. Simulated at $\lambda=1750$ nm.}
    
    \label{Fig4}
\end{figure}

A general reflective case occurs, illustrated in Figure~\ref{Fig4}(a), when the metasurface remains bianisotropic before and after reconfiguration ($\Delta A_1>0$, $\Delta A_2>0$). In this regime, the device exhibits both direction-dependent reflectivity and different reconfiguration thresholds under opposite illumination directions, resulting in an asymmetric reflective limiter. Since magnetoelectric coupling in State~1 simultaneously determines threshold separation $D$ and reflection contrast through Eq.~\eqref{eq: r}, increasing $\Delta A_1$ enlarges both quantities. Consequently, the reflectivity contrast and the transmission window can be jointly tailored through the degree of structural asymmetry. For the proposed metasurface, this regime is obtained at $\lambda=1400$~nm, where the electric and magnetic responses remain approximately balanced while preserving a finite magnetoelectric coupling in both states. 

%For the proposed metasurface, this regime is obtained at $\lambda=1400$~nm, where the electric and magnetic responses remain approximately balanced while preserving a finite magnetoelectric coupling in both states. Starting from our design, $D$ could be reduced by increasing the structural symmetry, for instance by reducing the VO$_2$ thickness or embedding the metasurface in a more symmetric dielectric environment, as discussed for the transmissive regime. In the limiting case, the structure would approach a reconfigurable mirror whose response is governed primarily by the PCM phase transition rather than by bianisotropy. Conversely, increasing the asymmetry, for instance by increasing the VO$_2$ thickness up to 95~nm, would further enhance both the reflectivity contrast and the threshold separation (see Supplementary Section S5).

A more extreme situation is obtained when the metasurface evolves from a strongly bianisotropic state to a nearly symmetric one ($\Delta A_1>0$, $\Delta A_2\approx0$), as illustrated in Figure~\ref{Fig4}(b). In this regime, the large absorption asymmetry of State~1 maximizes the threshold separation $D$, while the nearly vanishing asymmetry of State~2 suppresses the directional response after reconfiguration. As a result, the metasurface acts as a directional reflective limiter, exhibiting a power-dependent transition for only one illumination direction. For the proposed metasurface, this regime is approached at $\lambda=1490$~nm, where the design is close to the balanced-particle condition in State~1 while exhibiting minimal absorption asymmetry in State~2. Although the metallic VO$_2$ layer retains a finite broadband asymmetry, this functionality could be further enhanced through appropriate geometrical optimization. For example, incorporating a thin metallic layer above the silicon resonators would increase the asymmetry of State~1 while producing a more symmetric response after the VO$_2$ transition, thereby strengthening the directional limiting behavior.% (see Supplementary Section S??).

The opposite situation is obtained when a nearly symmetric state evolves into a strongly bianisotropic one ($\Delta A_1\approx0$, $\Delta A_2>0$), as illustrated in Figure~\ref{Fig4}(c). In this regime, the metasurface exhibits an almost identical response for both illumination directions at low optical intensities. Following the VO$_2$ transition, a pronounced reflection asymmetry emerges, resulting in a directional reflective switch. Unlike the previous case, the functionality is therefore determined primarily by the bianisotropy of State~2 rather than by the initial absorption asymmetry. 

For the proposed metasurface, this regime is obtained at $\lambda=1750$~nm. The directional switching contrast could be further enhanced by simultaneously reducing the asymmetry of State~1 and increasing that of State~2. This could be achieved by combining the design strategies discussed above, for example, by increasing the structural symmetry of the insulating state through dielectric cladding while reinforcing the asymmetry introduced after the VO$_2$ transition through a thicker active layer. Such an approach would enlarge $\Delta A_2$ while preserving $\Delta A_1\approx0$, maximizing the reconfigured reflection contrast.

\section{Conclusions} 

We have investigated the electromagnetic functionalities enabled by the thermal reconfiguration of bianisotropy in a silicon--VO$_2$ metasurface operating in the near infrared. By combining a collective-polarizability framework with numerical simulations and experimental characterization, we established a direct connection between the evolution of the electric, magnetic, and magnetoelectric responses and the resulting scattering properties. The experimental results confirm that the phase transition of VO$_2$ enables substantial modification of the magnetoelectric coupling, providing access to distinct bianisotropic states within a single metasurface platform.

The proposed framework reveals how different reconfiguration pathways of the absorption asymmetry give rise to distinct directional functionalities. In particular, we identify three representative reflective regimes corresponding to metasurfaces that remain bianisotropic after reconfiguration, evolve from asymmetric to symmetric responses, or exhibit the opposite transition. These regimes enable different forms of power-dependent asymmetric reflection, including directional limiting and switching functionalities, while a transmissive operating point is shown to support power-limiting transmission through asymmetric reconfiguration thresholds. Beyond the specific silicon--VO$_2$ implementation considered here, the presented approach provides general guidelines for engineering reconfigurable bianisotropic metasurfaces and highlights dynamic control of magnetoelectric coupling as a powerful degree of freedom for multifunctional and adaptive photonic devices.

Moreover, emissivity and absorptivity are equal, in this context, to the absorption of the metasurface. Therefore, the studied configurations could be used for passive thermal control, as the reconfiguration occurs at low temperatures and intensities. As shown, different direction-dependent emissivities can be engineered before and after the VO$_2$ phase change. These functionalities can be useful in radiative cooling in low-temperature applications~\cite{Radiative_cooling_2}.

\section{Methods}
\subsection{Numerical Simulations} 
The electromagnetic response of the metasurface was investigated using the Electromagnetic Waves, Frequency Domain module of COMSOL Multiphysics\textsuperscript{\textregistered}. The unit cell was modeled using Floquet periodic boundary conditions along the lateral directions, while periodic ports were implemented above and below the structure to excite normally incident plane waves and retrieve the corresponding scattering parameters. Unless otherwise stated, illumination was considered at normal incidence with the electric field polarized along the $x$ direction. The refractive indices of a-Si \cite{pierce1972electronic} and SiO$_2$ \cite{malitson1965interspecimen} were taken from literature, whereas the optical constants of VO$_2$ were obtained from spectroscopic ellipsometry measurements (Supplementary Section 2), and extrapolating linearly for the spectral range exceeding 2000 nm.

This simulation framework was employed both for eigenfrequency analyses and for the calculation of the scattering response of the metasurface. The eigenmode calculations were used to identify the resonant electric and magnetic dipolar modes and analyze their evolution during the VO$_2$ phase transition. Reflection and transmission coefficients were obtained from the simulated scattering matrix, from which the absorptance spectra were calculated. The collective electric, magnetic, and magnetoelectric polarizabilities were retrieved from the scattering parameters following a de-embedding procedure that removes the phase accumulated during wave propagation between the metasurface plane and the excitation ports. To ensure numerical convergence, the dimensions of the air domains and the position of the excitation ports were systematically varied. No significant variations were observed in the calculated eigenfrequencies, scattering parameters, or retrieved polarizabilities.

Thermally driven reconfiguration was modeled by coupling the Electromagnetic Waves and Heat Transfer in Solids modules. Thermal properties of VO$_2$, a-Si, and SiO$_2$ were taken from ~\cite{howes2020limiting}. Optical absorption within the structure was converted into a distributed heat source through the imaginary part of the complex refractive index of each material, resulting in a spatially resolved heating profile determined by the local electromagnetic field distribution. Consequently, the generated heat is not assumed to be uniform within the VO$_2$ layer, but directly follows the optical mode profiles supported by the metasurface.

The thermal simulations were performed using the same periodic boundary conditions adopted for the electromagnetic calculations, representing a locally periodic region of the illuminated metasurface, resulting in nearly identical optical excitation and thermal conditions for neighboring resonators. Heat dissipation through the substrate was modeled by setting the temperature on the bottom surface of the SiO$_2$ substrate to 25$^\circ$C. At the air interface, thermal exchange was described by a convective heat-flux boundary condition assuming ambient air at room temperature. 

Power-dependent functionalities were analyzed through steady-state thermal simulations for different incident optical intensities. For each excitation condition, the temperature distribution within the VO$_2$ layer was calculated and averaged over its bottom interface. The resulting temperature was then used to determine the corresponding phase state of the PCM and, consequently, its optical properties. By sweeping the incident intensity, the stationary optical response of the metasurface was obtained as a function of input power, enabling the identification of the transition thresholds associated with the different operating regimes presented in this work. The optical constants of VO$_2$ employed throughout the simulations were extracted from spectroscopic ellipsometry measurements performed on the fabricated films.

\subsection{Fabrication}

The VO$_x$ thin film was deposited onto a fused silica substrate by molecular beam epitaxy (MBE) and subsequently annealed to obtain the VO$_2$ phase. The growth procedure and annealing of the VO$_X$ films are described elsewhere\cite{van2015production}. The resulting film thickness was designed to be 77 nm. For metasurface fabrication, an amorphous silicon (a-Si) layer was deposited on top of the VO$_2$ film by plasma-enhanced chemical vapor deposition (PECVD). The metasurface pattern was defined by electron-beam lithography and transferred into the a-Si layer by inductively coupled plasma (ICP) dry etching, resulting in a periodic array of silicon nanodisks on the VO$_2$ layer. %Additional nanofabrication details can be found in Ref.~\cite{parra2021all}\textcolor{orange}{REVISAR, no tengo acceso al paper}.

\subsection{Characterization}

The crystalline structure of the VO$_2$ film was verified by grazing-incidence X-ray diffraction (GIXRD), as shown in Figure \ref{Fig1}(d). X-ray reflectometry (XRR) measurements were used to determine the film thickness and surface roughness, yielding values of 77 nm and 2.3 nm, respectively, from fitting of the experimental data. The optical constants of the VO$_2$ layer in the near-infrared spectral range were obtained by spectroscopic ellipsometry measurements performed at different temperatures and are provided in Supporting Information Section 2. Additionally, temperature-dependent electrical resistivity measurements were performed to verify the characteristic insulator-to-metal transition and hysteretic behavior of the VO$_2$ films. The corresponding resistivity curves are provided in Supporting Information Section 2. Following deposition of the amorphous silicon layer, its thickness was confirmed by spectroscopic ellipsometry, resulting in a value of 228 nm, corresponding to the height of the fabricated nanodisks.

The morphology of the metasurface was characterized by scanning electron microscopy (SEM) using secondary-electron detection with an accelerating voltage of 3 kV under low-current conditions to minimize sample charging. A representative SEM image of the fabricated structure is shown in Figure \ref{Fig3}(b).

The optical response of the metasurface was characterized by Fourier-transform infrared spectroscopy (FTIR) using a Vertex 80 spectrometer coupled to a Hyperion microscope. Measurements were performed using a custom objective providing near-normal illumination of the sample. Reflectance and transmittance spectra were acquired at room temperature and at 80$^\circ$C using an external heating stage consisting of a resistive heater with integrated thermistor (Thorlabs HT10KR1) controlled by a temperature controller (Thorlabs TC300). For reflectance measurements, an uncoated gold mirror was used as a reference, whereas transmission spectra were referenced to the optical path through the heating stage without the sample. The spectral responses were smoothed using a moving-average filter and provided in Supporting Information Section 3.

\section*{Acknowledgements}
The authors acknowledge the Spanish National Research Council (grant no. PID2024-162261NB-I00 and CNS2024-154715), the Universitat Polit\`ecnica de Val\`encia (grant no.PAID-01-23). This project has received funding from the European Union’s Horizon 2024 research and innovation program under the Marie Skłodowska-Curie grant agreement No 101226375

%\nolinenumbers
\printbibliography

\newpage
\section*{Supporting Information}
\renewcommand{\thefigure}{S\arabic{figure}}
\setcounter{section}{0} 
\setcounter{figure}{0}

\section{ Resonant modes analysis}
\begin{figure}[H] 
\centering 
\includegraphics[width=0.9\linewidth]{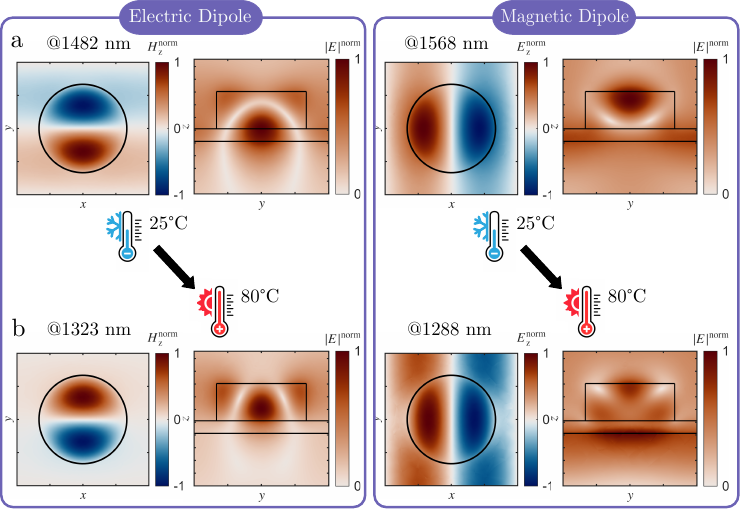} 
\caption{Normalized out-of-plane field components and electric-field amplitude distributions of the resonant eigenmodes supported by the metasurface. (a) Electric- and magnetic-dipole resonances for VO$_2$ in the insulating state (25$^\circ$C), located at $\lambda=$ 1482 nm and $\lambda=$ 1568 nm, respectively. (b) Electric- and magnetic-dipole resonances for VO$_2$ in the metallic state (80$^\circ$C), located at $\lambda=$ 1323 nm and $\lambda=$ 1288 nm, respectively. Despite the structural asymmetry of the metasurface, the modes can be predominantly identified as electric- and magnetic-dipole resonances.} \label{FigS1} 
\end{figure}

In Figure~\ref{FigS1}, the different field profiles for the two hybridize resonant modes are shown before and after the VO2 phase transition. The resonant wavelengths shift towards lower values due to the decrease of the VO2 refractive index in that range. 

\section{Characterization of VO$_2$ layer on SiO$_2$ substrate }
   
\begin{figure}[H] 
\centering 
\includegraphics[width=0.5\linewidth]{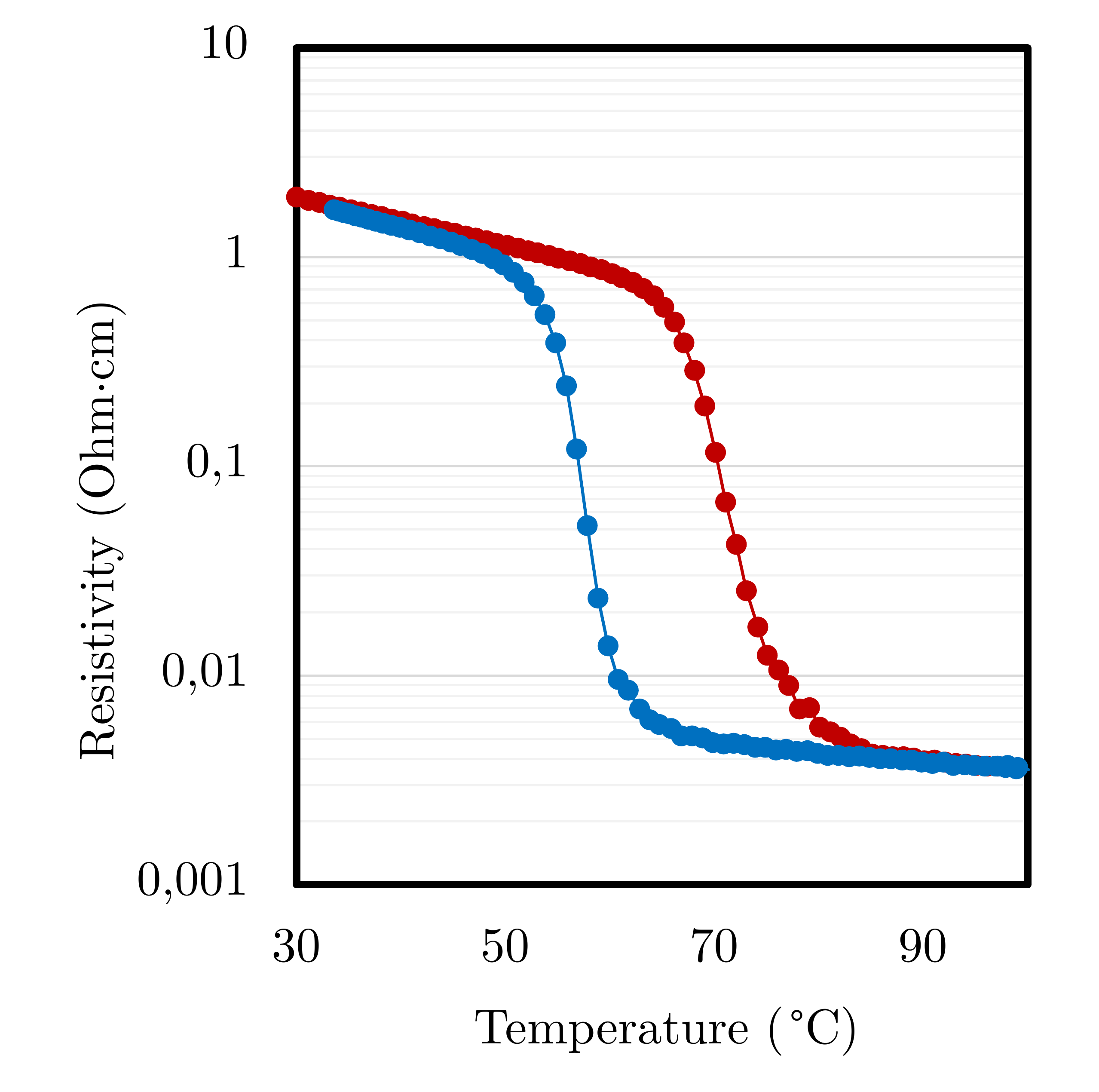} 
\caption{Temperature-dependent resistivity measurements of the VO$_2$ film, showing the characteristic hysteretic insulator-to-metal phase transition.} \label{FigS2} 
\end{figure} 
    
\begin{figure}[H] 
\centering
\includegraphics[width=0.43\linewidth]{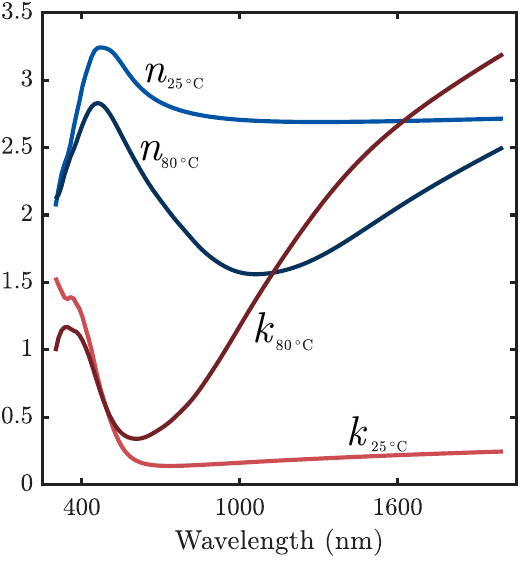} 
\caption{Complex refractive index, $n_{\rm c}=n-jk$, of the VO$_2$ film in its insulating and metallic states, extracted from spectroscopic ellipsometry measurements and used throughout the numerical simulations.} \label{FigS3} 
\end{figure}

\begin{figure}[H]
\centering 
\includegraphics[width=0.43\linewidth]{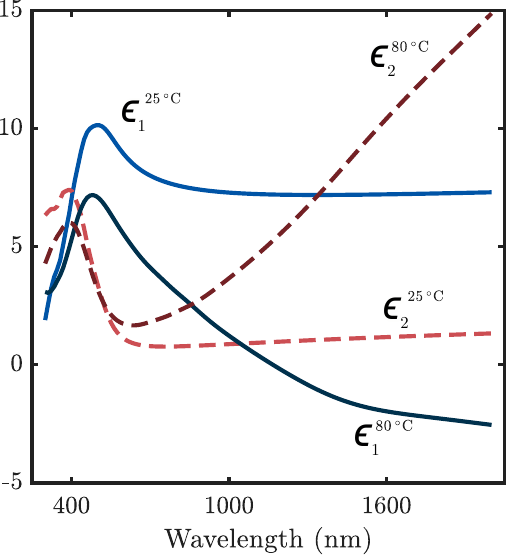}
\caption{Real and imaginary parts of the complex permittivity, $\varepsilon=\varepsilon_1-j\varepsilon_2$, of the VO$_2$ film in its insulating and metallic states, extracted from spectroscopic ellipsometry measurements. } \label{FigEps} 
\end{figure}

\section{Optical response of the fabricated metasurface}

The spectral responses were smoothed using a moving-average filter. Each data point was replaced by the arithmetic mean of its neighboring points within a sliding window of 100 samples (25 °C) or 20 samples (80 °C), reducing high-frequency fluctuations while preserving the overall spectral trend.
   
\begin{figure}[H] 
\centering
\includegraphics[width=0.43\linewidth]{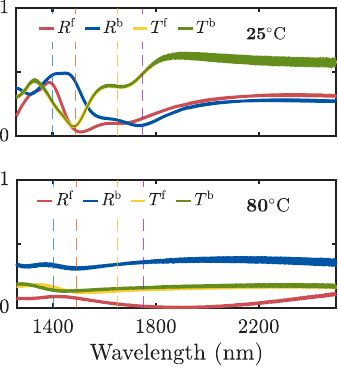} 
\caption{Experimental FTIR transmission and reflection spectra of the VO$_2$ sample at 25~$^\circ$C and 80~$^\circ$C, corresponding to the insulating and metallic phases, respectively. Dashed lines indicate the wavelengths ($1400$, $1490$, $1650$ and $1750$nm) at which the different functional regimes discussed in this work are obtained. For clarity, the spectra shown in the figure were smoothed as described in this section to suppress high-frequency fluctuations while preserving the overall spectral response.} \label{FigS4} 
\end{figure}

\begin{figure}[H]
\centering 
\includegraphics[width=0.43\linewidth]{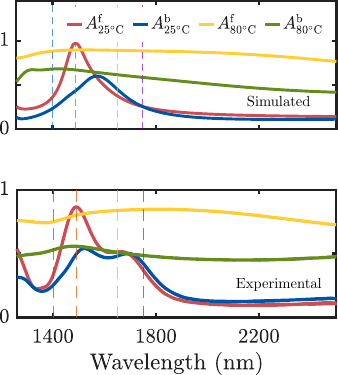}
\caption{Forward and backward absorption spectra of the VO$_2$ device at 25~$^\circ$C and 80~$^\circ$C obtained from numerical simulations (top) and FTIR measurements (bottom). The four absorption responses, $A^{\mathrm{f}}_{25^\circ\mathrm{C}}$, $A^{\mathrm{b}}_{25^\circ\mathrm{C}}$, $A^{\mathrm{f}}_{80^\circ\mathrm{C}}$, and $A^{\mathrm{b}}_{80^\circ\mathrm{C}}$, are shown separately to provide the individual contributions underlying the differential absorption $\Delta A$. Experimental spectra were smoothed using the procedure described in this section. } \label{FigS4} 
\end{figure}

\begin{figure}[H]
\centering 
\includegraphics[width=0.43\linewidth]{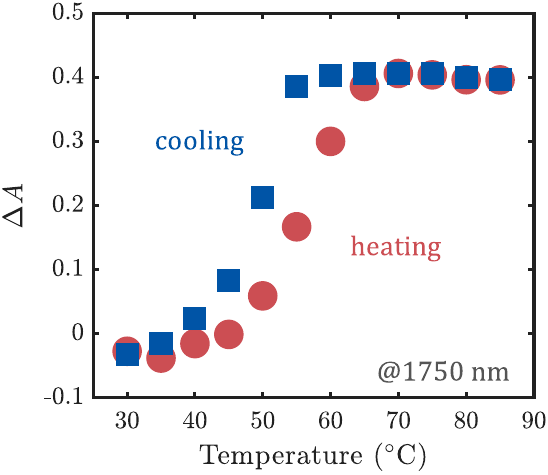}
\caption{Absorption asymmetry at a fixed wavelength of 1750 nm during a heating-cooling cycle, showing the experimentally observed hysteresis cycle of the reconfigurable electromagnetic response.  These measurements were performed on a metasurface designed with different geometrical parameters: $d=536$ nm, $P=800$ nm, $h=228$ nm. Nevertheless it demonstrates the role of the VO$_2$ transition in the reconfiguration of the bianisotropy.} \label{FigS4} 
\end{figure}

\section{ Complementary studies of asymmetric absorption}
    
\begin{figure}[H] 
\centering 
\includegraphics[width=0.53\linewidth]{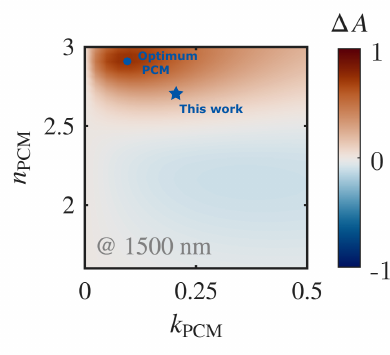} 
\caption{Absorption asymmetry at 1500 nm as a function of the real ($n$) and imaginary ($k$) parts of the PCM refractive index. The star indicates the optical constants of the VO$_2$ film characterized in this work at room temperature.} \label{FigS4} 
\end{figure}
    
\begin{figure}[H] 
\centering 
\includegraphics[width=0.43\linewidth]{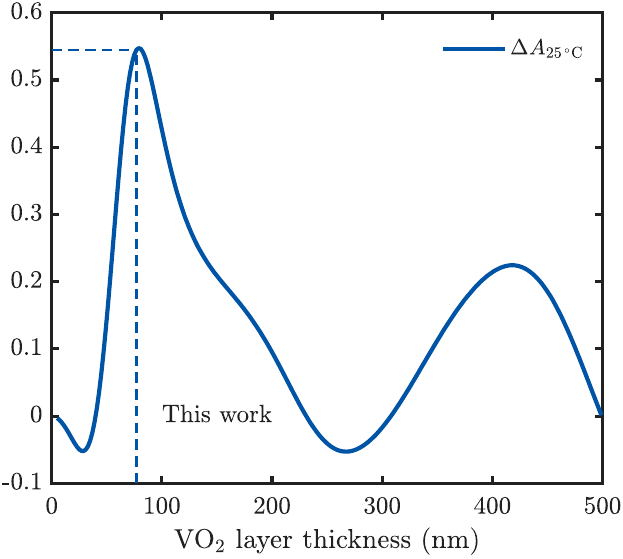} 
\caption{Absorption asymmetry at 1490 nm as a function of the VO$_2$ layer thickness. Increasing the VO$_2$ thickness enhances the structural asymmetry of the metasurface and enables larger absorption asymmetry values without modifying the remaining geometrical parameters.} \label{FigS5} 
\end{figure}

\begin{figure}[H] 
\centering 
\includegraphics[width=1\linewidth]{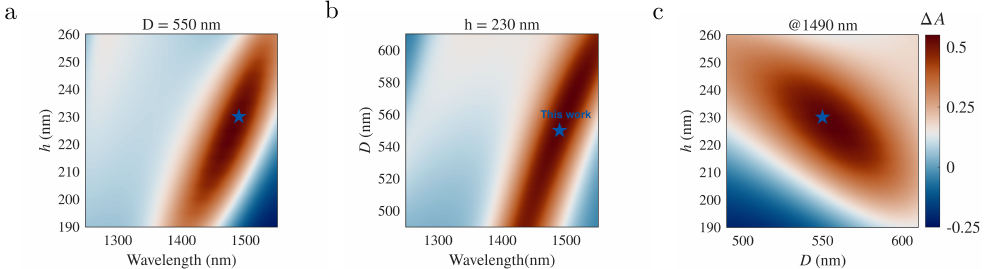} 
\caption{Absorption difference as a function of the dimensions, height and diameter, of the resonator and wavelength. In each case one of the variables is fixed, being: (a) $d=550$nm, (b) $h=230$nm and (c) wavelength at $1490$nm. The star marker indicates the values for the metasurface designed for this work. Geometrical parameters of the design address the region where the absorption difference is enhanced.   } \label{FigS5} 
\end{figure}

\end{document}